\documentclass[sigconf, authorversion, nonacm]{acmart}

\usepackage{multirow}

\AtBeginDocument{%
  \providecommand\BibTeX{{%
    \normalfont B\kern-0.5em{\scshape i\kern-0.25em b}\kern-0.8em\TeX}}}

\setcopyright{acmcopyright}
\copyrightyear{2018}
\acmYear{2018}
\acmDOI{10.1145/1122445.1122456}

\acmConference[CHI '21]{CHI '21: ACM conference on Human Factors in Computing Systems}{May 08--13, 2021}{Yokohama, Japan}

\begin{document}


\title{A Stretchable Electrostatic Tactile Surface}

\author{Naoto Takayanagi}
\email{naoto20221106@keio.jp}
\affiliation{
  \institution{Keio University}
  \city{Yokohama}
  \country{Japan}
}

\author{Naoji Matsuhisa}
\email{naoji@iis.u-tokyo.ac.jp}
\affiliation{
    \institution{Tokyo University}
    \city{Tokyo}
    \country{Japan}
}

\author{Yuki Hashimoto}
\email{hashimoto@iit.tsukuba.ac.jp}
\affiliation{%
  \institution{University of Tsukuba}
  \city{Tsukuba}
  \country{Japan}
}

\author{Yuta Sugiura}
\email{sugiura@keio.jp}
\affiliation{%
    \institution{Keio University}
    \city{Yokohama}
    \country{Japan}
}

\begin{abstract}
    Tactile sensation is essential for humans to recognize objects. Various devices have been developed in the past for tactile presentation by electrostatic force, which are easy to configure devices, but there is currently no such device that features stretchability. Considering that the device is worn over the joints of a human body or robot, it is extremely important that the device itself be stretchable. In this study, we propose a stretchable electrostatic tactile surface comprising a stretchable transparent electrode and a stretchable insulating film that can be stretched to a maximum of 50\%. This means that when attached to the human body, this surface can respond to the expansion and contraction that occur due to joint movements. This surface can also provide tactile information in response to deformation such as pushing and pulling. As a basic investigation, we measured the lower limit of voltage that can be perceived by changing the configuration of the surface and evaluated the states of stretching and contraction. We also investigated and modeled the relationship between the voltage and the perceived intensity.
\end{abstract}




\maketitle

\section{Introduction}
\label{Intro}
The sense of touch is indispensable for humans to recognize things, and research on tactile devices that present a sense of touch different from what is original there has been active in various fields. For example, in the field of mixed reality (MR), it is believed that tactile devices can provide users with a deeper sense of immersion\cite{Lu2021_chemical}\cite{yu2019skin}, and in the human interface field, tactile feedback is expected to improve user operability. 

Such tactile devices are mainly classified into those that provide mechanical stimulation, those that provide electrical stimulation, and those that modulate surface friction. Mechanically stimulated tactile devices utilize various types of actuators such as shape memory alloys, piezoelectric, and pneumatic actuators to directly deform the skin surface and provide tactile sensations. Electrically stimulated tactile devices present tactile sensations by directly stimulating nerves using electric currents\cite{Rahimi2019}. In tactile devices that modulate frictional force, the frictional force on the surface of an object is manipulated using electrostatic force or ultrasonic waves\cite{Ohmori2019}

While a wide variety of tactile devices have been developed for each of these approaches, it has been difficult to apply the interfaces developed so far attached to the skin surface of humans or soft robots.
This is because none of the interfaces have skin-like stretchability.
This has been particularly true when they have been attached over joints, or when they have been attached to the surface of soft objects and used in applications where large deformations occur, such as pushing or pulling.
In such cases, the tactile device itself needs to be flexible and stretchable.
In recent years, advances have been made in stretchable electronic materials, and in the field of tactile sensors, research is being conducted on a wide variety of sensors with stretchability\cite{jiang2020hyaline}\cite{jiang2021stretchable}\cite{yoon2022elongatable}.
We believe that these materials can be used to solve the problems of tactile devices.

\begin{figure}[!t]
	\centering
	\includegraphics[width=\linewidth]{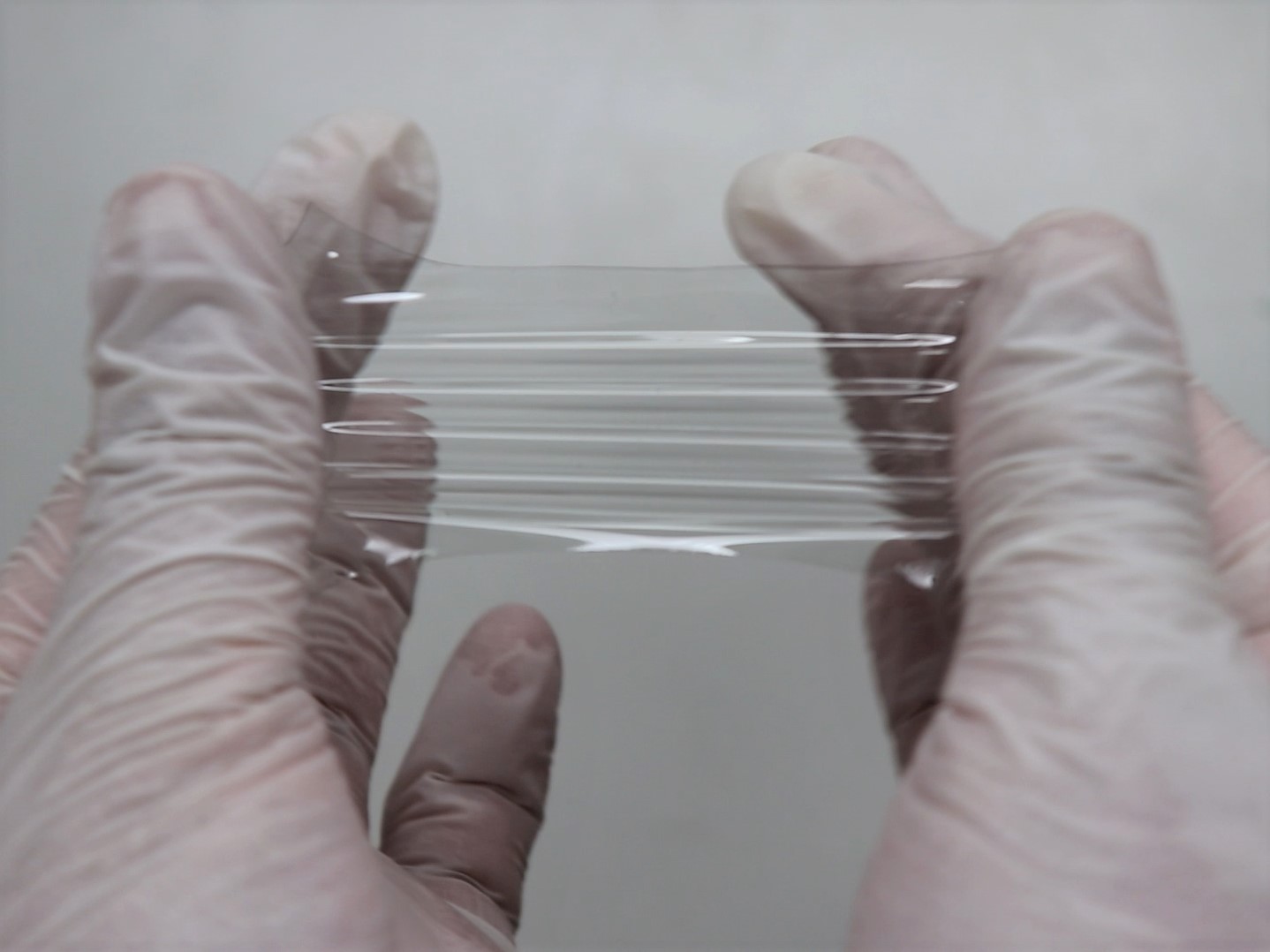}
\caption{Tactile surface proposed in this research as it is being stretched. The surface is sufficiently transparent and stretchable and can be applied on top of existing on-body input interfaces to add tactile feedback.}
	\label{fig:fig1}
\end{figure}

Therefore, our goal is to develop a stretchable tactile device (Fig. \ref{fig:fig1}). In this study, we focus our approach on the use of electrostatic tactile technology. An electrostatic tactile device requires only a conductive layer, an insulating layer, and a voltage source, which means it can be easily constructed. Ishizuka et al. proposed a tactile device that is flexible and can be attached to curved surfaces, although it is not stretchable because it does not use stretchable electronic materials\cite{ishizuka2018}. As mentioned above, it is not enough for the device to be flexible; it must also be stretchable, considering the joints of the human body and robots. We aimed to develop a stretchable tactile device by using a transparent stretchable electrode and a stretchable insulating layer. 

It is important to understand the range of voltages over which tactile sensations can be presented and how the intensity of the tactile sensations that can be presented varies with voltage when this surface is actually used. As a basic study for determining its practical use, we investigated the lower voltage limit of perception when changing the configuration of the surface and when changing the state of stretching. The relationship between the voltage and the perceived intensity was also investigated and modeled.

\section{Related work}
\label{RelatedWork}
\subsection{Electrostatic Tactile Display}
\label{ElectrostaticTactileDisplay}
\begin{figure}[!t]
	\centering
	\vspace{0.2in}
	\includegraphics[width=\linewidth]{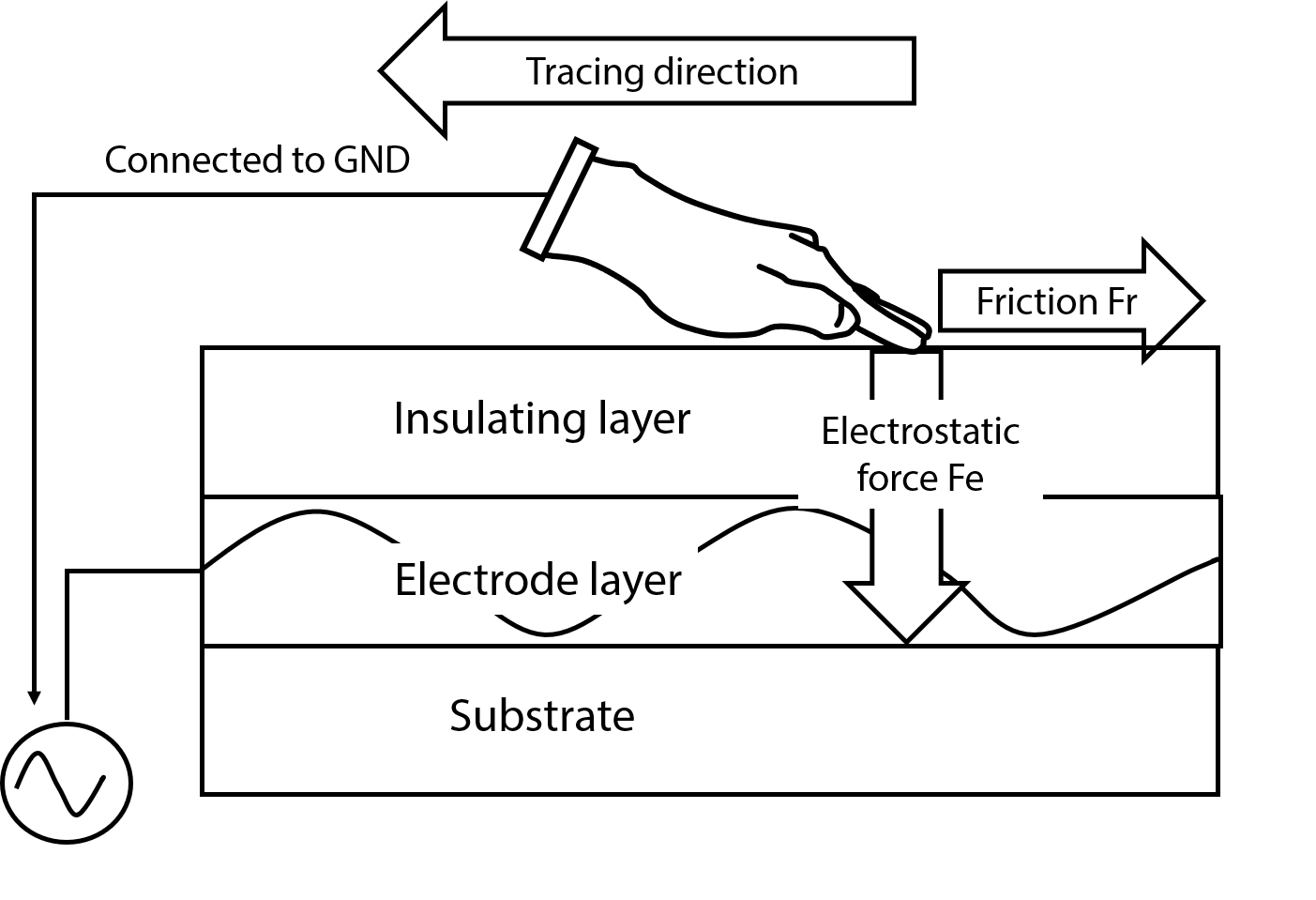}
\caption{Overview of electrostatic tactile technology. When the user traces a surface, electrostatic forces change the friction between the hand and the surface, resulting in tactile sensations.}
	\label{fig:display}
\end{figure}

An electrostatic tactile display is a device that provides tactile feedback utilizing electrostatic tactile technology. The principle is shown in Fig. \ref{fig:display}.
When an AC voltage is applied to the electrode, the hand is attracted to the display by the electrostatic force $F_e$. Here, $F_e$ is represented by Eq. (\ref{equ:1}), where $d$, $V$, $S$, and $\varepsilon$ denote the distance between the electrode and the fingertip, the applied voltage, the contact area, and the dielectric constant of the insulating layer, respectively.
\begin{equation}
F_e = \frac{{\varepsilon}SV^{2}}{2d^{2}}
\label{equ:1}
\end{equation}
The amount of $F_e$ varies with time and the frictional force $F_r$ is modulated accordingly,  providing the user with a different texture on the display surface (e.g., unevenness ). Since the time variation of the friction force can be controlled by the input waveform, a variety of textures can be provided simply by changing the waveform.

In an earlier study utilizing this technology, Bau et al. came up with a device that presents tactile sensation by sticking the device on a touch panel using transparent electrodes and an insulating layer, along with an AR system that provides tactile sensation to arbitrary objects by applying an electric current directly to the human body instead of to the object itself \cite{bau2010,bau2012}. Ishizuka et al. developed a high-resolution electrostatic haptic display using microfabricated electrode arrays \cite{ishizuka2017}. Both these studies demonstrated tactile sensations to the fingertips, and Fukuda et al. showed that it is possible to present tactile sensations to the palm as well \cite{fukuda2019basic}. With these prior works as an inspiration, we developed a new electrostatic tactile surface featuring elasticity by focusing on transparent stretchable electrodes and stretchable insulating films.

\subsection{Stretchable Sensors and Stretchable Electrodes}
\label{StretchableSensors}
Recent advances in materials science research have made it possible to fabricate highly stretchable sensors and displays. Weigel et al. developed a skin-worn sensor that can be stretched up to 30\% for touch input on the body \cite{weigel2015iskin}, and Wessely et al. developed a stretchable user interface that combines sensing capabilities and visual output \cite{wessely2016stretchis}. Xu et al. devised a stretchable keyboard \cite{xu2015stretch}. These studies utilized polydimethylsiloxane (PDMS), an organic polymer based on silicon that is easy to process. Since PDMS is both highly stretchable and insulating, we also use it as the insulating layer for the electrostatic tactile surface developed in the present study.

Poly(2,3-dihydrothieno-1,4-dioxin)-poly(styrenesulfonate) (PEDOT:PSS) is a typical conductive polymer that features excellent stability, transparency, and film formation properties compared to other conductive polymers, although it has low stretchability. Ishizuka et al. developed a flexible electrostatic tactile display using PEDOT:PSS sandwiched between PDMS substrates \cite{ishizuka2018}. Other studies have examined how to improve the stretchability of PEDOT:PSS. Wang et al. introduced an additive to PEDOT:PSS to improve its stretchability while maintaining its conductivity \cite{wang2017highly}. Using this as a reference, the conductive layer of our surface is composed of PEDOT:PSS with stretchability improved by additives.

\section{Implementation}
\subsection{Surface Construction}
\subsubsection{Materials}
\label{Materials}
Two types of material were utilized for the substrate of our surface: a styrenic thermoplastic elastomer (SEBS, Asahi Kasei Corporation, H1062) and a thermoplastic polyurethane (TPU, Japan Miractran Corporation, P22MBRNAT). For the conductive layer, we used PEDOT:PSS (Heraeus, PH1000) with added fluorosurfactant (Apollo Scientific Ltd., Capstone FS30) and lithium bis(trifluoro methanesulfonyl)imide (LiTFSI, Tokyo Chemical Industry Co., Ltd.). For the insulating layer, we used PDMS (Dow Inc., SILPOT 184). The prepolymer and curing agent were mixed at a weight ratio of 12:1.

\subsubsection{Fabrication Process}
\label{FabricationProcess}
The fabrication process of a surface for the experiment(50 mm $\times$75 mm) is described below.

First, SEBS (4 mL)  dissolved in cyclohexane at a concentration of 80 mg/mL was drop-casted onto a glass substrate. The SEBS was filtered through a glass fiber (GF) filter with a pore size of 1 $\mu$m to remove dust particles. After drop-casting, the samples were covered with a Petri dish and left overnight to dry. 

Next, to make the surface hydrophilic, it was treated with a UV ozone cleaning system (Novascan Technologies, Inc., Novascan PSD Pro-series) for 10 minutes. Then, TPU (1 mL) dissolved in DMF at a concentration of 100 mg/mL was spin coated using a spin coater (Hysol Corporation, ACE-200) at 1000 rpm for 60 seconds. After spin coating, the substrate was baked at 110 $^\circ$C for 10 minutes, and the surface was hydrophilized by UV ozone treatment for 10 minutes.

Next, we explain the conductive layer. LiTFSI (0.25 mL) dissolved in DIW at a concentration of 40 mg/mL and FS30 (10 $\mu$L) were added little by little to GF-filtered PEDOT:PSS (1 mL) before use, and mixed well. The resulting solution was spin coated onto the substrate at 1000 rpm for 60 seconds. After spin coating, the conductive layer was completed by baking at 110 $^\circ$C for 10 minutes.

\begin{figure}[tb]
	\centering
	\vspace{0.2in}
	\includegraphics[width=\linewidth]{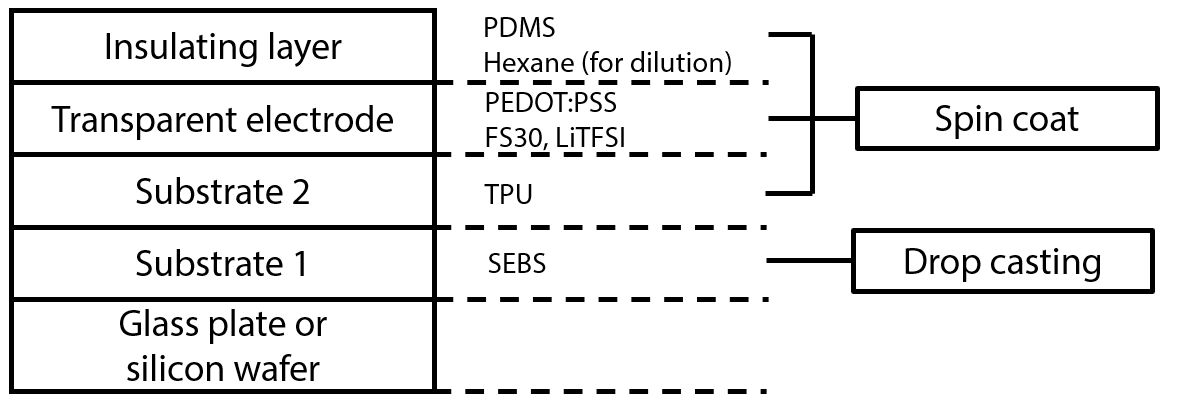}
\caption{Composition of stretchable electrostatic tactile surface consisting of four layers (substrate 1, 2, conductive layer, and insulating layer). The surface is made by stacking each layer in turn on top of a glass plate.}
	\label{fig:composition}
\end{figure}

Finally, we describe the insulating layer, which was composed of PDMS. We used a spinning and revolving mixer (Thinky Co., Ltd., ARE-310) to prevent the formation of air bubbles. Before spin coating the PDMS, solidified PDMS was attached to a part of the conductive layer to create a contact area connecting with the operating circuits. The PDMS mixture (approximately 3 g) was then spin coated onto the conductive layer at 1000 rpm for 60 seconds. After that, the insulating layer was annealed at 120 $^\circ$C for 30 minutes, and then peeled off from the glass substrate to complete the stretchable electrostatic tactile surface. The composition of this surface is shown in Fig. \ref{fig:composition}.

The influence of the insulating layer thickness on the lower perceptible voltage was investigated by diluting the PDMS solution with hexane. The thickness of the film can be adjusted according to the dilution ratio. Using this method, we fabricated three types of electrostatic tactile surface with insulating layers of different thicknesses. We measured each insulating layer with a laser microscope (KEYENCE, VK-X1000) and confirmed that the thickness was 57.5 $\mu$m (undiluted solution), 17.9 $\mu$m (PDMS:hexane=2:1 weight ratio mixture), and 7.2 $\mu$m (PDMS:hexane=1:2 weight ratio mixture).

\subsection{Equipment Configuration}
\label{EquipmentConfiguration}
\begin{figure}[tb]
	\centering
	\vspace{0.2in}
	\includegraphics[width=\linewidth]{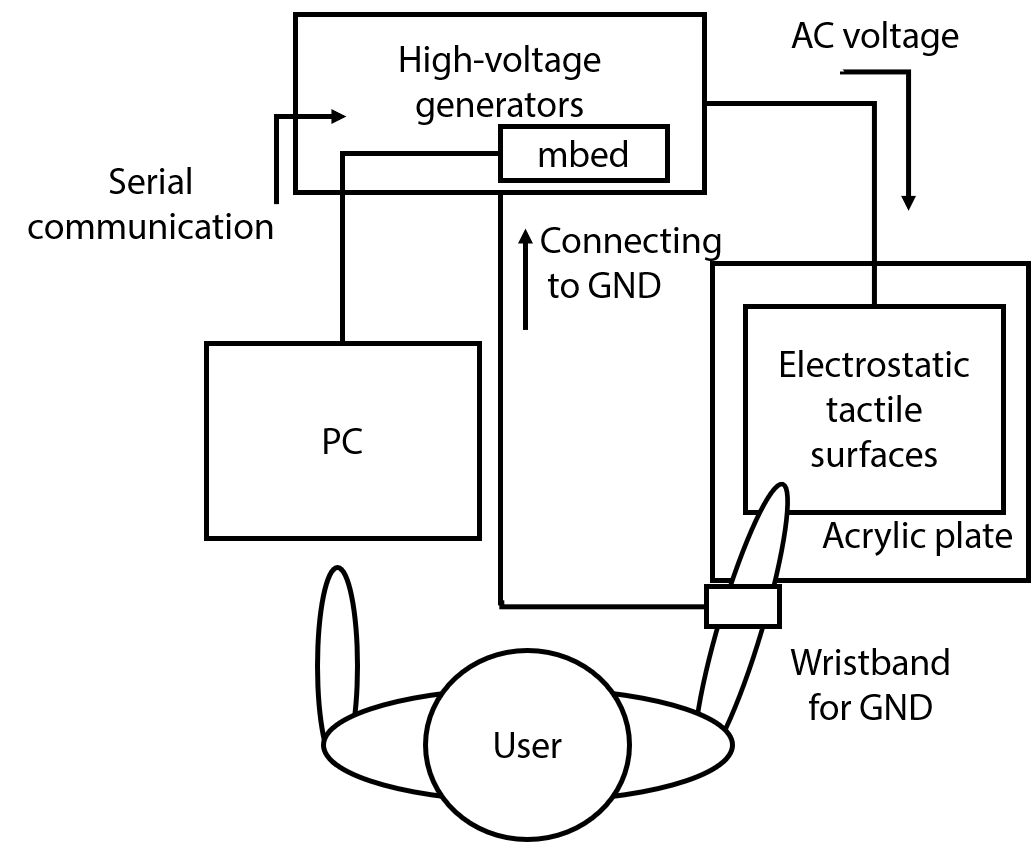}
\caption{Configuration of apparatus in Experiments 1--3. The voltage value and frequency of the AC voltage applied to tactile surfaces are controlled from a PC via mbed. The user wears a wristband for grounding.}
	\label{fig:configuration}
\end{figure}

The configuration of the apparatus used in the experiments is shown in Fig. \ref{fig:configuration}.
The electrostatic tactile surface was fixed to an acrylic plate. A PC and a high-voltage generator were connected by a USB cable via mbed (NXP, LPC1768).

The output terminal of the high-voltage generator was connected to the electrode of the surface, and the GND terminal was connected to a GND wristband. We constructed the high-voltage generator ourselves with reference to the literature \cite{yem2016comparative}. The maximum applied voltage of the generator is 570 V. In this device, the current is controlled to flow only up to 0.6 mA, which has no effect on the human body. Its frequency can also be adjusted, but in the following experiments, we fixed the frequency to a sinusoidal wave of 100 Hz, which previous studies have indicated has a high perceived intensity \cite{pyo2014new}.

\section{Experiments 1, 2: Lower Perceivable Voltage Limit Investigation}
\label{EX1,2}
We investigated the perceivable voltage to determine the voltage at which this surface can present tactile sensations to the user. Measurements were made in accordance with the limit method. Participants traced the surface an arbitrary number of times with and without voltage applied and were asked to choose whether or not there was a difference in tactile perception between the two conditions. If participants felt there was a difference, it indicated they perceived electrostatic tactile sensation, and if they felt no difference, it meant no electrostatic tactile sensation was perceived. 

First, the voltage was lowered in 50 V increments from the maximum output voltage of 570 V, and participants were asked if they perceived a difference. The voltage value at which each participant finally indicated there was a difference was $V_x$. The voltage was measured by changing the voltage in 10 V increments, with $V_x - 100$ as the reference value during the ascent and $V_x + 50$ as the reference value during the descent. This was done to shorten the experimental time. The voltage at which the difference was first felt was recorded during the ascent, and the voltage at which the difference was last felt was recorded during the descent. Three sets of these processes were performed for each condition. During the experiment, participants were instructed to wipe their hands with tissue paper to prevent tactile attenuation due to sweaty hands.

\subsection{Experiment 1: Relationship between insulating layer thickness and lower perceptible voltage}
\label{EX1}
\subsubsection{Purpose}
As described in \ref{ElectrostaticTactileDisplay}, the amount of electrostatic force received by the participant is inversely proportional to the distance between the electrode and the fingertip, or the square of the thickness of the insulating layer (Eq. (\ref{equ:1})). We felt that reducing the thickness of the insulating layer would increase the sensitivity, so we investigated the relationship between the thickness of the insulating layer and the lower perceptible voltage limit using three surfaces with different thicknesses.

\subsubsection{Conditions}
In Experiment 1, we utilized all of the surfaces listed in 3.A.2. For this experiment, we used a surface the size of a glass plate. The participants were instructed to trace the surface with three fingers. The participants were four males between the ages of 21 and 23 (mean 22.3 years, standard deviation 0.83 years). The measurement took approximately 45 minutes per participant. In total, 4 (participants) × 3 (surface types) × 2 (series) × 3 (sets) = 72 (times) measurements were made.

\subsubsection{Results}
\begin{figure}[tb]
	\centering
	\vspace{0.2in}
	\includegraphics[width=\linewidth]{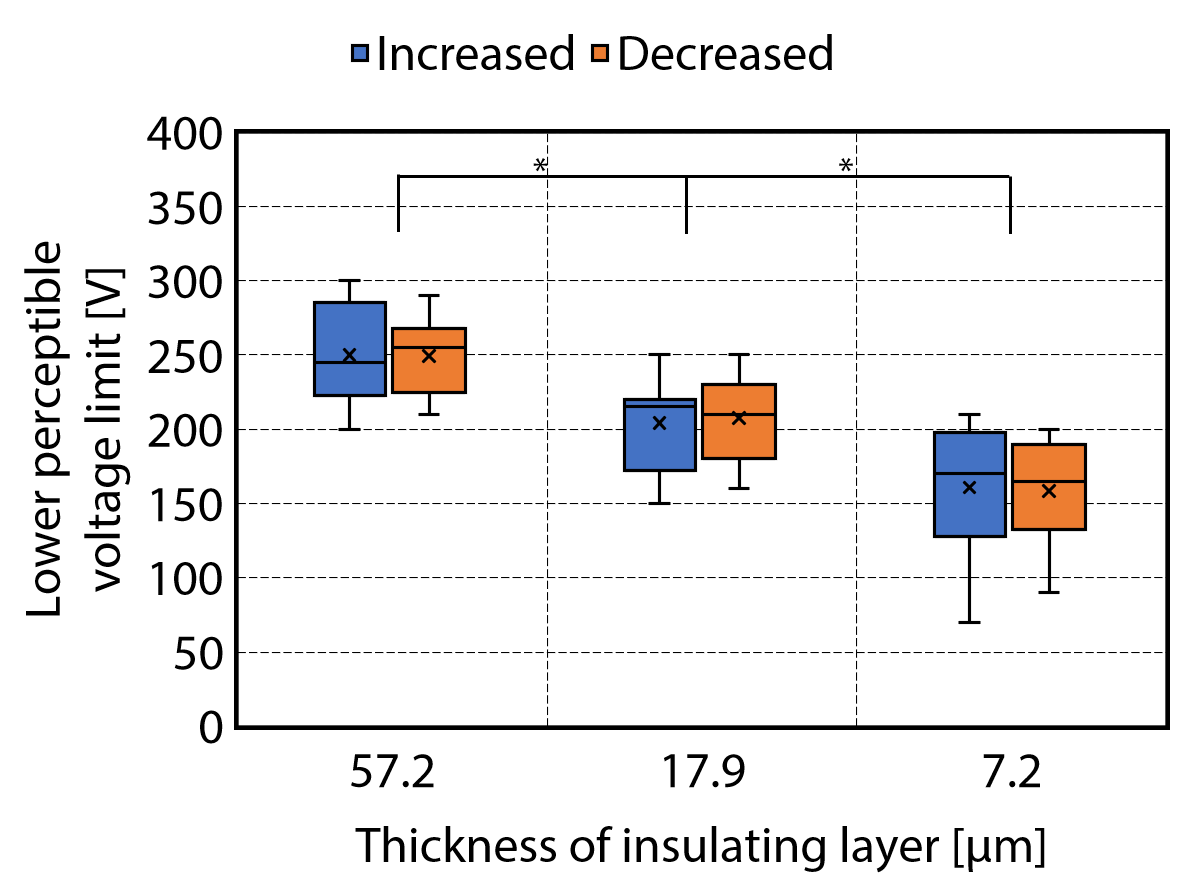}
\caption{Influence of insulation layer thickness on lower perceptible voltage limit}
	\label{fig:ex1}
\end{figure}

Box-and-whisker plots of lower perceptible voltage recorded in the ascending and descending series for each surface are shown in Fig. \ref{fig:ex1}.
Multiple comparison tests using the Kruskal-Wallis test and the Dunn-Bonferroni method showed a significant difference between the three surfaces  (p\textless0.05). This result indicates that the sensitivity could be improved by making the insulating layer thinner.

\subsection{Experiment 2: Relationship between stretching state and lower perceptible voltage}
\label{EX2}
\subsubsection{Purpose}
The most important feature of our proposed tactile surface is that it is stretchable. Therefore, in Experiment 2, we measured the lower perceptible voltage limit when the surface was stretched or retracted and determined whether it was possible to use the surface with stretching or retraction.

\subsubsection{Conditions}
In Experiment 2, we utilized the surface with an insulating layer thickness of 57.5 $\mu$m. Four stretching states were implemented: 25\% stretching, 50\% stretching, 50\% stretching and then back to the original state, and no stretching. We used double-sided tape and regular tape to fix the surface from above and below, respectively, in order to fix the stretching state of the surface. The participants were instructed to trace the surface with one finger. The participants were four males between the ages of 21 and 23 (mean 22.0 years, standard deviation 1.00 years). The measurement took approximately 30 minutes per participant. In total, 4 (participants) × 4 (stretch conditions) × 2 (series) × 3 (sets) = 96 (times) measurements were made.

\subsubsection{Results}
\begin{figure}[tb]
	\centering
	\vspace{0.2in}
	\includegraphics[width=\linewidth]{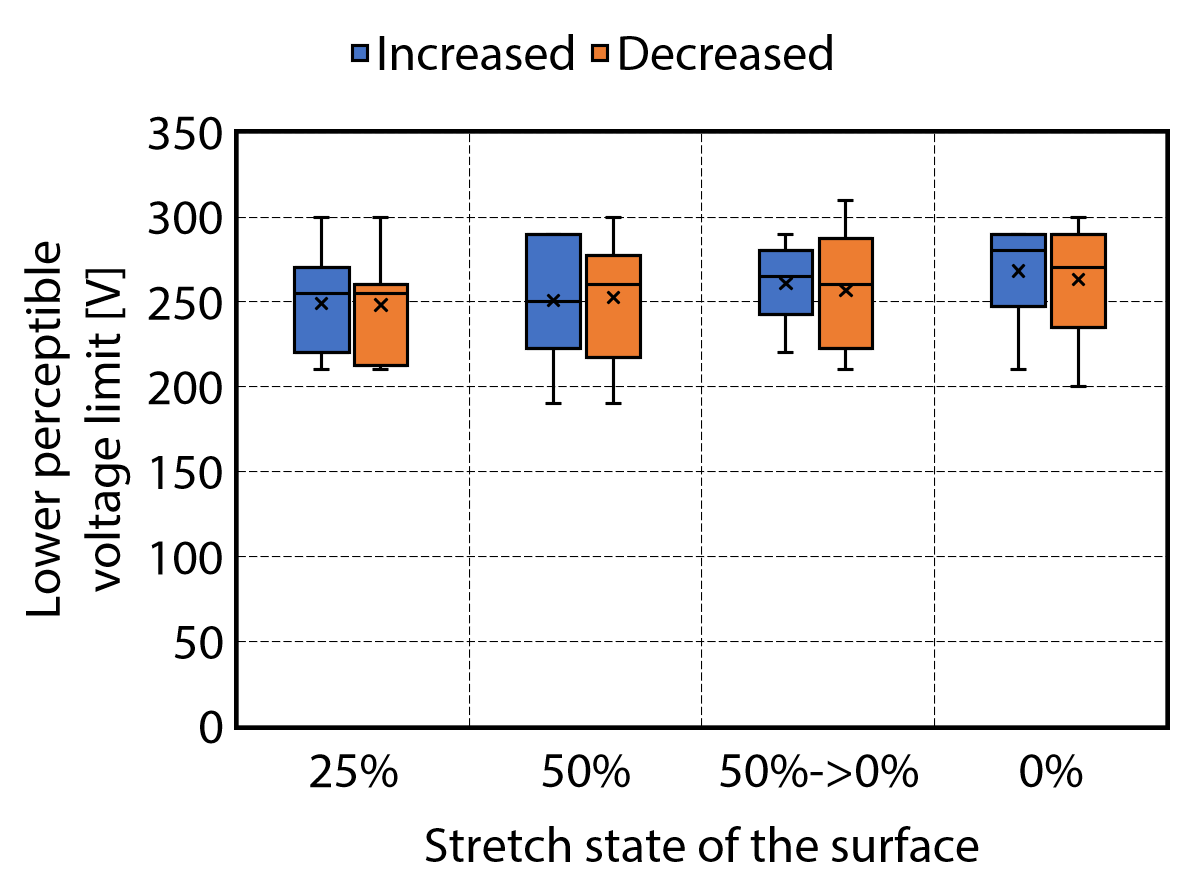}
\caption{Influence of stretch state of surface on lower perceptible voltage limit.}
	\label{fig:ex2}
\end{figure}

Box-and-whisker plots of the lower perceptible voltages recorded in the ascending and descending series for each state are shown in Fig. \ref{fig:ex2}.

Multiple comparison tests using the Kruskal-Wallis test and the Dunn-Bonferroni method were conducted, but there were no significant differences. This indicates that the electrostatic tactile surface can be extended and retracted without any loss of sensitivity.

\section{Experiment 3: Relationship between Voltage and Perceived Intensity Ratio}
\label{EX3}
By examining the lower perceptible voltage limit in \ref{EX1,2}, we were able to determine the voltage at which the surface can operate. When constructing a system using this surface, it is important to know not only the range of the operating voltage of the surface but also the extent to which the perceived intensity varies. In this section, we experimentally investigate and model the relationship between the voltage and the perceived intensity ratio.

\subsection{Conditions}
Based on the results of \ref{EX1,2}, we used  360 V as the reference voltage at which the user can stably perceive tactile sensations. Four different voltages (360 V, 430 V, 500 V, and 570 V) from the base voltage to the maximum applied voltage were utilized in the experiment. These voltages were presented randomly and the perceived intensity was compared. The participants were seven males aged between 21 and 25 years (mean 23.6 years, standard deviation 1.02 years), and the experiment took approximately 45 minutes per participant.

\subsection{Procedure}
Participants first traced the surface with their index finger while a voltage of 360 V was applied, and then memorized the perceived intensity. Using this as a reference, they continued to trace the surface until they were sure that they could remember it. Next, the voltage was randomly selected from four values (360 V, 430 V, 500 V, and 570 V) and participants were asked to indicate how many times the perceived intensity increased relative to the reference. Participants were instructed to answer with any number greater than 0. If they were not confident in their answer the first time, they were allowed to trace the surface again as many times as they wished. If they forgot the reference voltage, they returned to the reference voltage and continued the experiment. Participants checked the perceived intensity at the reference voltage (360 V) each time they responded. This process was conducted 8 times for each voltage value, for a total of 4 (voltage types) × 8 (sets) × 3 (surface types) = 96 times. The participants were instructed to wipe their hands with tissue paper to prevent tactile attenuation.

\subsection{Results}
\begin{figure}[tb]
	\centering
	\vspace{0.2in}
	\includegraphics[width=\linewidth]{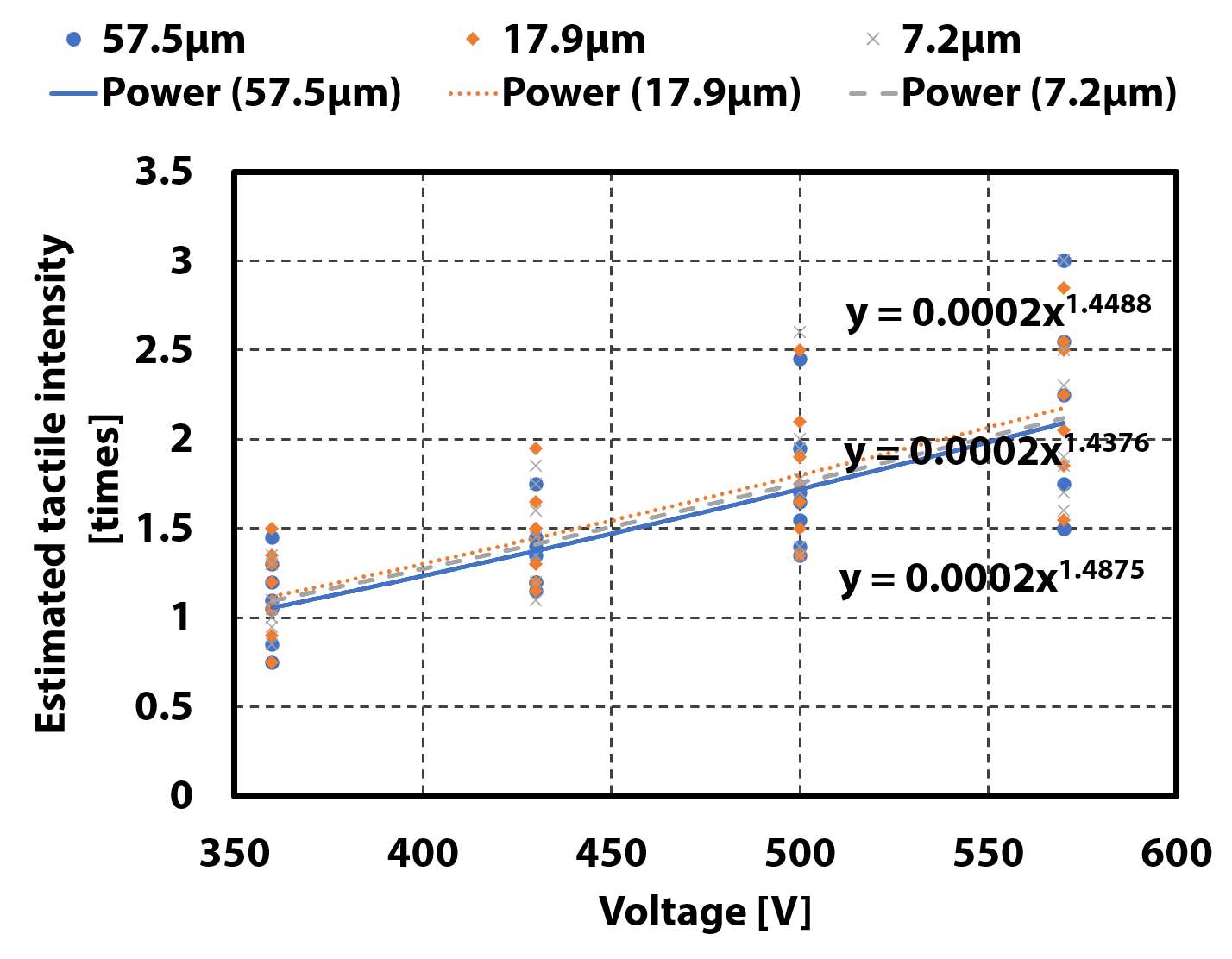}
\caption{Relationship between voltage and perceived intensity ratio.}
	\label{fig:ex3}
\end{figure}

A graph plotting each participant’s median response to voltage values and surface types is shown in Fig. \ref{fig:ex3}.

The graph also shows the approximate curve obtained on the basis of Stevens' power law. This law states that the following equation can be obtained empirically for a physical stimulus quantity $S$ and a psychological quantity $\phi$ using constants $k$ and $n$ that vary depending on the type of stimulus.

\begin{equation}
\label{equ:2}
\phi = kS^n 
\end{equation}

In this experiment, we obtained the power-law approximation curve with the voltage value as the stimulus amount $S$ and the participant’s response as the perceived intensity $\phi$. The constants $k$ and $n$ obtained for each surface were $k=0.0002$ and $1.4875$ for surfaces with an insulating layer thickness of $57.5\,\mathrm{\mu m}$, $k=0.0002$ and $n=1.4488$ for surfaces with an insulating layer thickness of 17.9 $\mu$m, and $k=0.0002$, $n=1.4376$ for a surface with an insulating layer thickness of 7.2 $\mu$m.

Multiple comparison tests were performed using the Kruskal-Wallis test and the Dunn-Bonferroni method. When the voltage values were fixed, there was no significant difference in the perceived intensity ratio between the surface types. In contrast, when the surface type was fixed, there were significant differences in perceived intensity ratios between all voltage values except between 500 V and 570 V for surfaces with an insulation layer of 7.2 $\mu$m (p\textless0.05). This suggests that the thickness of the insulation layer has little effect on the change in perceived intensity and that the rate of change in perceived intensity is determined by the change in voltage. As for the reason, there was no significant difference in the perceived intensity ratios between 500 V and 570 V for surfaces with an insulation layer of 7.2 $\mu$m, although we did not precisely examine this difference in the experiment, several participants said that the surface with a smaller insulation layer thickness had stronger perceived intensity at the same voltage. As discussed in 4.A.2, this is presumably because the thickness of the insulating layer increases the electrostatic force. It is generally believed that the larger the stimulus amount, the more difficult it is to accurately perceive changes in the stimulus. It is possible that the surface with an insulating layer of 7.2 $\mu$m had greater perceived intensity at 500 V and 570 V than the other surfaces, which is why the changes were not accurately perceived.

\section{Application}
\label{App}
\subsection{On-Body Interface}
\label{On-Body}
\begin{figure}[tb]
	\centering
	\vspace{0.2in}
	\includegraphics[width=\linewidth]{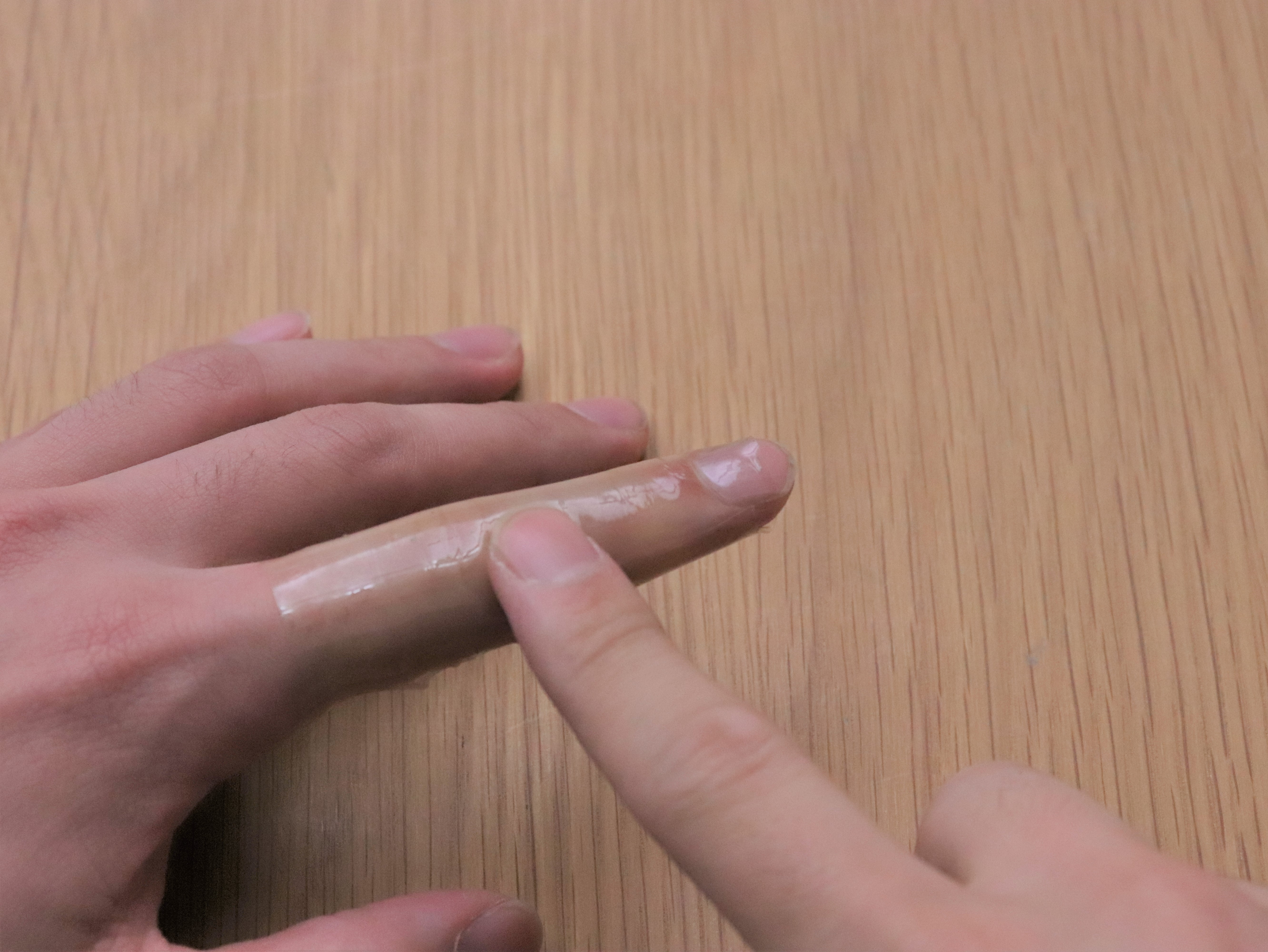}
\caption{Image of On-Body Interface. The stretchability of the surface allows it to be used without problems even when it is attached to jointed parts such as fingers and arms.}
	\label{fig:On-Body}
\end{figure}

\begin{figure}[tb]
	\centering
	\vspace{0.2in}
	\includegraphics[width=\linewidth]{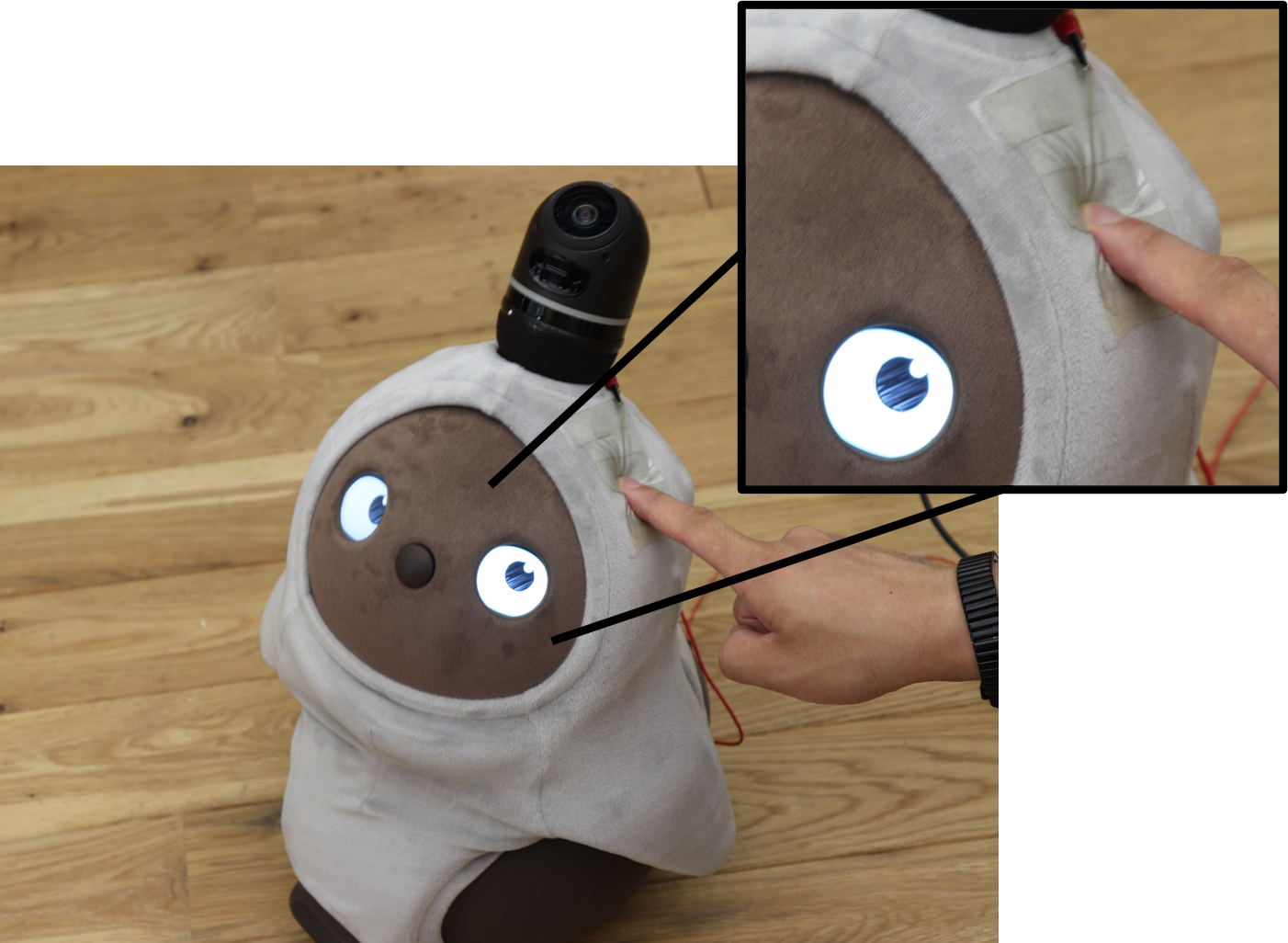}
\caption{Image of On-Body Interface attached to a soft robot. The stretchability of this surface allows it to withstand deformation by being pushed into soft objects. The surface is transparent enough that it does not spoil the appearance.}
	\label{fig:Lovot}
\end{figure}
Recent research has been conducted on using the arm or other parts of the body as an input interface or display by attaching a device to the surface of the human skin. In addition to the skin-mounted sensors introduced earlier\cite{weigel2015iskin}, there are other input interfaces such as the thin and flexible multi-touch sensors developed by Nittala et al.\cite{Nittala}. Other studies provide visual feedback at the same time as accepting input\cite{Weigel2017}. The surface in our study is very thin and is transparent enough to add tactile feedback by attaching it over the devices proposed in these other works. Its stretchability will presumably enable it to be attached over joints as well (Fig. \ref{fig:On-Body}).

\subsection{Touch Control of Soft Robots}
\label{SoftRobot}
Recent research attention has focused on soft robots, which behave differently depending on the emotion they wish to convey. For example, they might wag their tails and purr when they are happy, or shake their heads and cry when they are displeased. Real-life creatures respond in the same way, but in addition, their sense of touch on the surface of their bodies can change in response to their emotions. For example, humans get goosebumps when they feel fear, and animals such as dogs and cats raise their hair when they threaten their enemies. Therefore, we feel that the quality of interaction can be improved by changing the sense of touch depending on the state and situation of a soft robot (Fig. \ref{fig:Lovot}). Of course, thanks to the stretchability of the surface, it can be applied over a robot with many joints without any problem, just as it is applied to the human body.

\section{Discussion and Limitations}
\label{Discussion}
The results of Experiment 1 demonstrate that a thinner insulating layer on the surface increases the frictional force and lowers the perceivable lower limit voltage. On the other hand, in Experiment 2, there was no significant difference in the lower limit of perceivable voltage when the surface was stretched by 50\%. This seems to contradict the idea that stretching the surface makes the insulating layer thinner. In Experiment 2, the insulating layer was subjected to a maximum strain of 50\% which, by simple calculation, should have reduced the thickness of the insulating layer by two-thirds. However, since the thickness of the insulating layer between each surface used in Experiment 1 changed by about one-third, we can assume that Experiment 2 did not produce enough change in the insulation layer to make a significant difference. As a result, as long as the surfaces are used with up to 50\% stretch, the change in the perceivable lower voltage limit due to stretching does not need to be considered.

Although we were able to develop a stretchable electrostatic tactile surface in this study, it has some limitations. First, the stretchability is limited to 50\%. The material of each layer used for this surface can theoretically be stretched up to 100\%, but we found that the insulating layer peels off when touched with a stretch of 75\% to 100\%. It is possible that the affinity between the insulating layer and the conductive layer used in this study was low. This could be improved if the surface of PEDOT:PSS could be modified. Improving the stretch ratio is an issue to be addressed in the future.

Next, there is the problem of only a short period of usability. The insulating layer often broke during use, especially when the insulating layer was thinner. On the other hand, although we have not conducted a specific experiment on this, the thinner the insulating layer, the stronger the perceived strength, so there is a trade-off between perceived strength and durability. Improving the durability and investigating the relationship between the thickness of the insulation layer and the ratio of perceived intensity will be the focus of future work. 
In each experiment, there were only four to seven male participants aged 21 and 25, and it is unclear whether the same results would be obtained for women or for users in other age groups. In addition, this experiment was conducted on hard surfaces, namely, glass and acrylic plates. While we confirmed that the tactile sensation itself can be obtained when the device is attached to a soft surface, which is the intended use of the device, we have not yet conducted a precise investigation of the lower voltage limit and the perceived intensity ratio. Further experiments are needed.

\section{Conclusion}
\label{Conclusion}
In this study, we utilized transparent stretchable electrodes and stretchable insulating layer thin films to construct a stretchable electrostatic tactile surface. Our findings showed that sensitivity can be improved by making the insulating layer thinner. The surface was able to be used in a range of elasticity up to 50\% without loss of sensitivity. We also investigated the relationship between voltage and perceived intensity ratio and created a model of the rate of change in perceived intensity for this surface. In future work, we will improve the stretchability and durability, expand the range of experimental participants, and investigate the behavior of the surface when applied to soft surfaces.

\begin{acks}
This research was partially supported by JST PRESTO (JPMJPR20B7), JST AIP-PRISM (JPMJCR18Y2), and the Foundation for the Promotion of Ion Engineering.
\end{acks}

\bibliographystyle{ACM-Reference-Format}
\bibliography{reference.bib}

\appendix








\end{document}